\documentclass{ws-mplb}
\usepackage{ulem}
\usepackage{graphicx}
\usepackage{booktabs}
\usepackage{epstopdf}

\begin{document}

\markboth{Y. \it{Wu}, H. \it{Chang}, G. \it{Guo} \& S. \it{Lin}}{Multi-party quantum key agreement protocol  with authentication}
\catchline{}{}{}{}{}
%
%%%%%%%%%%%%%%%%%%%%%%%%%%%%%%%%%%%%%%%%%%%%%%%%%%%%%%%%%%%%%%%%%%%%

\title{Multi-party quantum key agreement  protocol  with authentication}

\author{Yiting Wu}

\address{College of Mathematics and Information,\\ Fujian Normal University,\\ Fuzhou 350007, China}

\author{Hong Chang}

\address{College of Mathematics and Information,\\ Fujian Normal University,\\ Fuzhou 350007, China}

\author{Gongde Guo}

\address{College of Mathematics and Information,\\ Fujian Normal University,\\ Fuzhou 350007, China}

\author{Song Lin}

\address{College of Mathematics and Information,\\ Fujian Normal University,\\ Fuzhou 350007, China\\ lins95@gmail.com}

\maketitle
\begin{history}
\received{(Day Month Year)}
\revised{(Day Month Year)}
\end{history}

\begin{abstract}
Utilizing the advantage of quantum entanglement swapping, a multi-party quantum key agreement protocol with authentication is proposed. In this protocol, a semi-trusted third party is introduced, who prepares Bell states, and sends one particle to multiple participants respectively. After that the participants can share a Greenberger-Horne-Zeilinger state by entanglement swapping. Finally, these participants measure the particles in their hands and obtain an agreement key.  Here, classical hash function and Hadamard operation are utilized to authenticate the identity of participants. The correlations of GHZ states ensure the security of the proposed protocol. To illustrated it detailly, the security of this protocol against common attacks is analyzed, which shows that the proposed protocol is secure in theory.
\end{abstract}

\keywords{Quantum key agreement; quantum entanglement swapping; authentication.}
\section{Introduction}

Unlike classical cryptography whose theoretical basis is computational complexity, the security of quantum cryptography is ensured by the principles of quantum mechanics. Since quantum cryptography is unconditionally secure in theory, it is used to solve some secure tasks, and forms some research branches, such as quantum key distribution(QKD),\cite{1,2,3}  quantum secret sharing(QSS),\cite{4,5,6}  quantum private comparison(QPC),\cite{7,8,9,10}  quantum secure direct communication (QSDC), \cite{11,12,13,14,15}  quantum key agreement(QKA),\cite{16,17,18,19,20,21,22,23,24,25,26,27,28,29,30} etc. Key agreement (KA) is an important cryptographic primitive that is widely used in access control, key generation, and so on. \cite{31} In a key agreement, two or more users agree on a key in such a way that they equally influence the negotiated key. That is, the key cannot be determined by any subset of the participants. Moreover, the generated key is private and shared by all users. In 2004, Zhou {\it et al.}\cite{16} proposed the first quantum key agreement protocol. In this protocol, two users utilize the entanglement of EPR pairs to share a secret key. In 2013, Shi {\it et al.}\cite{17} studied the situation of multiple participants and proposed the first multiparty quantum key agreement protocol based on entanglement swapping. Subsequently, quite a few multiple quantum key agreement protocols are presented, which make full use of various characteristics of quantum mechanics.

In practical applications, there exists a case, in which one external attacker impersonates an internal participant and executes the key agreement protocol. Evidently, this attacker can successively eavesdrop the agreement key at the end of the protocol. Hence, identity authentication needs to be considered in designing a quantum key agreement protocol, which is the same as quantum cryptography protocols.\cite{32,33,34,35,36,37} However, almost all existing QKA protocols do not take it into consideration and are insecure to the impersonation attack. In this paper, an authenticated multi-party quantum key agreement protocol based on entanglement swapping is proposed. In this protocol, a semi-trusted third party (called TP) is introduced to generate Bell states and transmits these signal particles to the participants. Here, entanglement properties of Bell states are utilized to authenticate the identity of each participant. Then, by entanglement swapping, all participants share a GHZ state that is used to generate the agreement key. Meanwhile, the correlations of GHZ states is utilized to detect eavesdropping, which can ensure the security of the protocol. Finally, these participants achieve key agreement task by measuring the particles of GHZ states.

The rest of the paper is organized as follows. In the next section, some fundamental preliminaries are introduced. In Sect. 3, an authenticated three-party QKA protocol is proposed. The security analysis of this protocol is provided in Sect. 4, and the generalization to the multiple parties is in Sect. 5. The paper ends with Sect. 6 where conclusions are drawn.
\section{Preliminaries}

In this protocol, EPR pairs are used as the signal carriers, which are in one of the four Bell states:
\begin{eqnarray}
|\phi^{\pm}\rangle=\frac{1}{\sqrt{2}}(|00\rangle\pm|11\rangle),\ |\psi^{\pm}\rangle=\frac{1}{\sqrt{2}}(|01\rangle\pm|10\rangle),
\end{eqnarray}
where $B_z=\{|0\rangle, |1\rangle\}$ is an orthogonal basis of a two-dimensional Hilbert space, called $Z$-basis. Moreover, $B_x=\{|\pm\rangle=\frac{1}{\sqrt{2}}(|0\rangle\pm|1\rangle)\}$ forms another set of orthogonal basis, called $X$-basis. Hadamard operator $H=\frac{1}{\sqrt{2}}(|0\rangle\langle0|+|0\rangle\langle1|+|1\rangle\langle0|-|1\rangle\langle1|)$ can realize the interchange between these two groups of bases, i.e., $H|0\rangle=|+\rangle$, $H|1\rangle=|-\rangle$. In addition, if we perform operation $H\otimes H$ on an EPR pair, the quantum state remains unchanged, i.e., $H\otimes H|\phi^{\pm}\rangle=|\phi^{\pm}\rangle ,\ H\otimes H|\psi^{\pm}\rangle=|\psi^{\pm}\rangle$. However, if we only perform $H$ operation on one particle, the quantum state becomes the superposition of two Bell states,
\begin{eqnarray}
I\otimes H |\phi^{\pm}\rangle=\frac{1}{\sqrt{2}}(|\phi^{\mp}\rangle+|\psi^{\pm}\rangle),
\nonumber\\
I\otimes H |\psi^{\pm}\rangle=\frac{1}{\sqrt{2}}(|\phi^{\pm}\rangle-|\psi^{\mp}\rangle),
\end{eqnarray}
where $I$ is the identity matrix.

Besides, GHZ state is another common entangled state. The three-particle GHZ state can be expressed as follow,
\begin{eqnarray}
|\varphi_{abc}\rangle=\frac{1}{\sqrt{2}}(|0bc\rangle+(-1)^{a}|1\rangle|b\oplus1\rangle|c\oplus1\rangle),
\end{eqnarray}
where, $a,b,c=\{0,1\}$, and the symbol $\oplus$ denotes addition modulo 2. By performing the following corresponding Pauli operations
\begin{eqnarray}
I=|0\rangle\langle0|+|1\rangle\langle1|,\ X=|0\rangle\langle1|+|1\rangle\langle0|,
\nonumber\\
Z=|0\rangle\langle0|-|1\rangle\langle1|,\ \mathrm{i}Y=|0\rangle\langle1|-|1\rangle\langle0|,
\end{eqnarray}
on the particles, $|\varphi_{000}\rangle$ can be converted to the state $|\varphi_{abc}\rangle$. The details are as follows,
\begin{eqnarray}
|\varphi_{abc}\rangle=Z^{a}\otimes X^{b}\otimes X^{c}|\varphi_{000}\rangle,
\end{eqnarray}
where $Z^{0}=X^{0}=I$.

Entanglement swapping is a nice property of quantum mechanics, which is widely used in the field of quantum information. Suppose there are three EPR pairs, (1,2), (3,4), (5,6), which are all in the state $|\phi^{+}\rangle$, i.e., $|\phi^{+}\rangle_{12} \otimes |\phi^{+}\rangle_{34} \otimes |\phi^{+}\rangle_{56}$. After performing GHZ states measurements on particles 1, 3, 5, the remaining three particles will collapse to a corresponding GHZ state. It can be expressed as
\begin{eqnarray}
|\phi^{+}\rangle_{12}\otimes|\phi^{+}\rangle_{34}\otimes|\phi^{+}\rangle_{56}=\frac{1}{2\sqrt{2}}\sum_{a,b,c=o}^{1}|\varphi_{abc}\rangle_{135}|\varphi_{abc}\rangle_{246},
\end{eqnarray}
From the equation above, it's obvious that if the measurement result is $|\varphi_{abc}\rangle$,the other three particles also collapse to $|\varphi_{abc}\rangle$. This conclusion will be used to design the proposed protocol.

\section{Three-party QKA protocol}

For the sake of simplicity, we first describe the three-party QKA protocol, then generalize it to a multi-party case in later section.
In this protocol, three participants P$_1$, P$_2$, P$_3$ wish to jointly negotiate a secret key $K$, the length of which is $n$. Meanwhile, they hope to authenticate the identity of the participants with the assistance of a semi-trusted third party TP. That is to say, TP is allowed to misbehave on his own, but cannot conspire with any party. Beforehand, each participant P$_i(i=1,2,3)$ has his own identity information ID$_i$ and shares a private key $k_i$ with TP, where ID$_i$ is public, $k_i$ is private. As is shown in Fig.1, these participants can achieve this task via the following steps.

\begin{figure}
\begin{center}
\includegraphics[width=5 in]{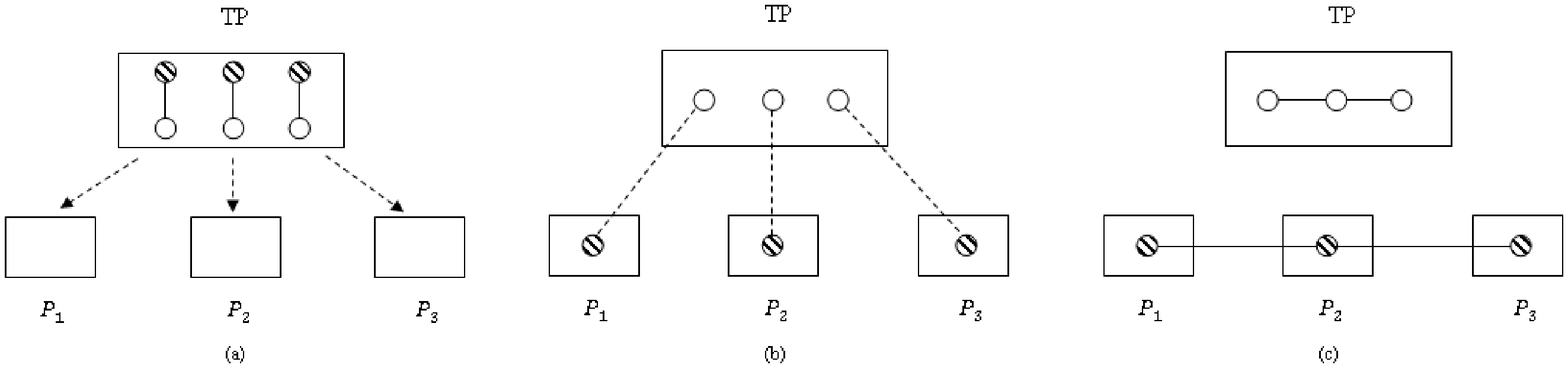}
\end{center}
\caption{\label{fig:one} Three-party quantum key agreement protocol.}
\end{figure}

\noindent Step 1 \begin{minipage}[t]{0.9\linewidth}TP and participant P$_i(i=1,2,3)$ separately generate a random number $r_T$ and $r_i$, and declare them publicly. Then, TP randomly selects a hash function $h$ from a hash function cluster and announces it, where $h:2^{*}\rightarrow2^{l}$, and $l=n+3\delta$. Each participant  P$_i$ and TP calculate the hash value $t_i$, where $t_i=h_{k_i}(ID_i||r_i||r_T)$.
\end{minipage}

\noindent Step 2 \begin{minipage}[t]{0.9\linewidth}TP prepares 3 ordered sequences of $l$ Bell states. Each EPR pairs is in the initial state $|\phi^{+}\rangle$. TP uses first qubits of each Bell state to form 3 ordered sequences $Q_1,Q_2,Q_3$. Similarly, he forms another 3 ordered sequences $S_1,S_2,S_3$ with all the second qubits. Here, $H_{i}$, $S_{i}(i=1,2,3)$ denote the first and second particles of the Bell state $|\phi^{+}\rangle$. Then, TP generates $3\zeta$ decoy particles which are randomly in four states $\{|0\rangle,|1\rangle,|+\rangle,|-\rangle\}$. He respectively inserts $\zeta$ decoy particles randomly in sequences $S_1,S_2,S_3$, and obtains 3 new particle sequences $S_1^{'},S_2^{'},S_3^{'}$. Finally, TP sends $S_1^{'},S_2^{'},S_3^{'}$ to P$_1$, P$_2$, P$_3$ respectively (see Fig.1(a)).
\end{minipage}

\noindent Step 3 \begin{minipage}[t]{0.9\linewidth}(The first eavesdropping detection) Confirming that P$_i$ has received the particle sequence $S_i^{'}$, they begin to check the eavesdrop. For sequence $S_{i}^{'}$, TP announces the positions and the corresponding basis ($B_{z}$ or $B_{x}$) of decoy particles. Then, P$_i$ measures the decoy particles in correct basis and randomly announces half of the measurement results. TP announces the initial states of the left half of decoy particles. At last, they check whether the initial states and the measurement results are consistent. If they are consistent, TP and P$_i$ think the quantum channel between them is secure; otherwise, they abandon the protocol.
\end{minipage}

\noindent Step 4 \begin{minipage}[t]{0.9\linewidth}Both P$_i$ and TP perform operation $I$ or $H$ respectively on two particles of one Bell state according to the value of $t_i$. Concretely, if $t_i=0$, they perform operation $I$; otherwise, they perform operation $H$. Obviously, if the selected operations of P$_i$ and TP are the same, the state of each EPR pairs stays unchanged. Otherwise, the Bell states will be in the form as Eq.(2) (see Fig.1(b)).
\end{minipage}

\noindent Step 5 \begin{minipage}[t]{0.9\linewidth}TP sequentially makes GHZ states measurements on 3 particles in his hand, and announces the measurement result $|\varphi_{abc}\rangle$. According to Eq.(6), the corresponding particles in three participants' hands will collapse to $|\varphi_{abc}\rangle$ (see Fig.1(c)).
\end{minipage}

\noindent Step 6 \begin{minipage}[t]{0.9\linewidth}P$_i$ selects an appropriate operation from $\{I,X,Z\}$ according to the result announced by TP. Specifically, when the result is $|\varphi_{abc}\rangle$, P$_1$, P$_2$ and P$_3$ performs $Z^{a}$, $X^{b}$ and $X^{c}$ on their particles respectively, where $Z^{a}\otimes X^{b}\otimes X^{c}|\varphi_{abc}\rangle=|\varphi_{000}\rangle$. In this way, three participants can share $l$ GHZ entangled states, all of which are in the state $|\varphi_{000}\rangle$.
\end{minipage}

\noindent Step 7 \begin{minipage}[t]{0.9\linewidth}(The second eavesdropping detection) P$_1$ randomly selects $\delta$ particles from the particle sequence $S_{1}$ for eavesdropping detection, and informs the other two participants P$_2$, P$_3$ of the positions of detecting particles. For each detecting particle, P$_1$ randomly measures it in basis $B_{z}$ or $B_{x}$. Then, he requires P$_2$ and P$_3$ to measure their qubits in the same basis as his, respectively. After that, P$_2$ and P$_3$ tell P$_1$ their measurement outcomes. Suppose that the measurement outcomes of three participants in the basis $B_{z}$ are $j_{P_i}^z=\{+1,-1\}$, and that in $B_{x}$ are $j_{P_i}^x=\{+1,-1\}$. According to
\begin{eqnarray}
|\varphi_{000}\rangle&=&\frac{1}{\sqrt{2}}(|0\rangle|0\rangle|0\rangle+|1\rangle|1\rangle|1\rangle)
\nonumber\\
&=&\frac{1}{2\sqrt{2}}(|+\rangle|+\rangle|+\rangle+|+\rangle|-\rangle|-\rangle+|-\rangle|+\rangle|-\rangle+|-\rangle|-\rangle|+\rangle),
\end{eqnarray}
P$_1$ checks whether or not $j_{P_1}^z=j_{P_2}^z=j_{P_3}^z$ is satisfied for those measurements in $B_{z}$(whether $\prod_{i=1}^{3}j_{P_i}^x=1$ is satisfied or not in $B_{x}$). If the error rate exceeds the threshold, they abondon the protocol. Otherwise, they continue the protocol.
\end{minipage}

\noindent Step 8 \begin{minipage}[t]{0.9\linewidth}P$_2$ and P$_3$ execute the similar eavesdropping detection as P$_1$ does in Step 7. Obviously, the detecting particles that P$_2$ and P$_3$ select do not have a common element with that of P$_{1}$.
\end{minipage}

\noindent Step 9 \begin{minipage}[t]{0.9\linewidth}Each participant measures his remaining particles in the basis $B_{z}$ and can get the raw agreement key $K$.
\end{minipage}

\section{Security Analysis}
The security of the above protocol will be discussed in this section. In a QKA protocol, one participant may be dishonest, thus both the external and internal attack should be considered. In addition, it is necessary to discuss the impersonation attack because this proposed protocol is designed to authenticate the identity of participants. In the following, the security of the presented protocol under these three kinds of attacks is analyzed.

\subsection{External Attack}
In this kind of attack, the purpose of an external attacker Eve is to eavesdrop the final negotiated key $K$ or the private key $k_{i}$. The public messages in this protocol are the identity of each participant, i.e., ID$_{i}(i=1,2,3)$, the random numbers $r_{T}$ and $r_{i}$, the hash function $h$, and the measurement results of TP. Obviously, if Eve does not attack the particle sequences, she can't obtain $K$ or $k_{i}$ according to the above public messages.

Since the particles are transmitted only once in this protocol, Eve has to attack the transmission of particle sequences $S_{i}^{'}$ in Step 2. However, the security of transmissions of the signal particles is based on decoy states, which has been proved to be secure. In other words, any of Eve's attack will be detected in the first eavesdropping check process. Hence, the proposed protocol is secure against external attacks.

\subsection{Internal Attack}
\subsubsection{Dishonest participants' attack}
The case, in which there is only one dishonest participant, is considered firstly. The goal of this dishonest participant is to determine the negotiated key $K$ alone, which breaks the fairness of the presented protocol. Without loss of generality, we can assume that P$_{1}$ is a dishonest participant, denoted as $\bar{\rm P}_{1}$. After Step 8, the particles in each participant's hands are in maximally mixed state. If $\bar{\rm P}_{1}$ wants the $d$-th bit in the negotiated key $K$ to be 0, he measures the $d$-th bit of the remaining particles in the $B_{z}$ basis. The measurement result will randomly be $|0\rangle$ or $|1\rangle$.  If it is $|0\rangle$ ($|1\rangle$), the measurement results of other participants are also $|0\rangle$ ($|1\rangle$). It shows that the $d$-th bit of the negotiated key is randomly 0 or 1. Namely, $\bar{\rm P}_{1}$ cannot determine any bit of the negotiated key. So, in order to achieve his goal, $\bar{\rm P}_{1}$ has to attack the transmitted particles. However, it is the same as external attacks in the previous section. That is, his attack will be detected in the first eavesdropping check process. Thus, one dishonest participant can't break the fairness of the proposed protocol.

Besides, there exists another common attack strategy, the collusive attack, in which two or more dishonest participants cooperate to attack the protocol. In general, it is more powerful than the attack mentioned above. In Step 9, the corresponding remaining particles are GHZ states. So no matter who measures his particles firstly in the basis $B_{z}$, the measurement results of three participants are the same. That is to say, all the participants fairly influence the negotiated key. It shows that the collusive attack is the same as the previous attack in this presented protocol. Therefore, the collusion of dishonest participants cannot undermine the fairness of the protocol.

Another case, in which $\bar{\rm P}_{1}$ wants to eavesdrop the private key $k_{i}$ of an honest participant P$_{i}$, should be considered. If $\bar{\rm P}_{1}$ wishes to eavesdrop $k_{i}$, he has to attack the transmitting particle sequence $S_{i}^{'}$. Since there only requires the participation of TP and participants in the first eavesdropping detection process. Attacking the particle sequence $S_{i}^{'}$ makes $\bar{\rm P}_{1}$ be detected as the external attacker Eve. Therefore, it is inevitable that $\bar{\rm P}_{1}$'s attack has to introduce errors if he wants to eavesdrop $k_{i}$.

\subsubsection{TP's attack}
In this protocol, TP is not required to be honest. So, TP's attack should be considered. Moreover, TP is semi-trusted. In this case, the purpose of TP is to eavesdrop the negotiated key $K$ without breaking the protocol. It's clear that the first eavesdropping detection is invalid to TP. In order to obtain the negotiated key $K$, TP prepares special fake particles instead of $|\phi^{+}\rangle$.

For example, if TP prepares $|00\rangle$, then the whole system state is in the state,
\begin{eqnarray}
|\Psi\rangle&=&|0\rangle_{T_{1}}|0\rangle_{P_{1}}\otimes |0\rangle_{T_{2}}|0\rangle_{P_{2}}\otimes |0\rangle_{T_{3}}|0\rangle_{P_{3}}
\nonumber\\
&=&\frac{1}{2}\sum_{a=0}^{1}|\varphi_{a00}\rangle_{T_{1}T_{2}T_{3}}(|\varphi_{a00}\rangle+|\varphi_{a\oplus1,0,0}\rangle)_{P_{1}P_{2}P_{3}},
\end{eqnarray}
Obviously, after TP's measurements, the particles in three participants' hands will collapse to $|\varphi_{a00}\rangle$ or $|\varphi_{a\oplus1,0,0}\rangle$. It is easy for TP to deduce every bit of the negotiated key $K$, which is either 0 or 1. Obviously, if the particles in three participants' hands are $|\varphi_{a00}\rangle$, there are no errors to occur. However, if the particles in their hands are $|\varphi_{a\oplus1,0,0}\rangle$, TP's attack will be detected in the second eavesdropping detection. Concretely, three participants performs $Z^{a}\otimes X^{0}\otimes X^{0}$ on $|\varphi_{a\oplus1,0,0}\rangle$ respectively in Step 6, then the whole quantum system is in the state $|\varphi_{100}\rangle$. Obviously, it will introduce errors when P$_{i}$ measures the detecting particles in the basis $B_{x}$. Hence, TP's attack will be detected with the probability of $(1-(\frac{1}{4})^\delta)$ in the second eavesdropping detection.

Next we consider a more general case. TP prepares $|\alpha\rangle=|000\rangle|\theta_{0}\rangle+|111\rangle|\theta_{1}\rangle$ instead of $|\phi^{+}\rangle$. TP sends the particle sequences to P$_{i}(i=1,2,3)$ and retains the ancilla. At the end of the protocol, TP tries to get the negotiated key $K$ by observing the ancilla without being detected. When P$_{i}$ measures the detecting particles in the basis $B_{z}$, TP's attack evidently does not introduce errors in the second eavesdropping detection. In order that no errors introduced in the basis $B_{x}$, $|\alpha\rangle$ needs to satisfy the following conditions.
\begin{eqnarray}
\langle++-|\alpha\rangle=\langle+-+|\alpha\rangle=\langle-++|\alpha\rangle=\langle---|\alpha\rangle=0,
\end{eqnarray}
From the Eq.(9), we can deduce $|\theta_{0}\rangle=|\theta_{1}\rangle$. Therefore, we can rewrite $|\alpha\rangle$ as
\begin{eqnarray}
|\alpha\rangle=(|000\rangle+|111\rangle)|\theta_{0}\rangle,
\end{eqnarray}
From the above equation, it is evident that $|\alpha\rangle$ is a product of a GHZ state $|\varphi_{000}\rangle$ and the ancilla. This implies that TP cannot gain any information about $K$ from observing the ancilla. Consequently, the proposed protocol is secure against TP's attack.

\subsection{Impersonation Attack}
In practical applications, there exists a case, in which one external attacker impersonates an internal participant and executes the key agreement protocol. The goal of an impersonal participant is to obtain the final agreement key $K$. Without loss of generality, suppose participant P$_{1}$ is an impersonal participant, denoted as $\hat{\rm P}_{1}$. Since $\hat{\rm P}_{1}$ does not know the real $k_{1}$, he can't calculate the hash value $t_1=h_{k_1}(ID_1||r_1||r_T)$. Therefore, he has to randomly select unitary operations $\{I,H\}$ in Step 4. If the selected operation is the same as TP's, it does not introduce any error. Otherwise, his attack will be detected in the second eavesdropping detection. The detailed analysis is depicted as follows.

Suppose that $\hat{\rm P}_{1}$ chooses a different operation from TP's. That is, he performs $I\otimes H$ on the Bell state $|\phi^{+}\rangle$, i.e., $I\otimes H|\phi^{+}\rangle_{T_{1}\hat{P}_{1}}=\frac{1}{2}(|00\rangle+|01\rangle+|10\rangle-|11\rangle)_{T_{1}\hat{P}_{1}}$. After TP's measurements, the whole state will change from Eq.(6) to
\begin{eqnarray}
&&(I\otimes H) |\phi^{+}\rangle_{T_{1}\hat{P}_{1}}\otimes |\phi^{+}\rangle_{T_{2}P_{2}}\otimes |\phi^{+}\rangle_{T_{3}P_{3}}\nonumber\\
=&&\frac{1}{4}\sum_{a,b,c=0}^{1}|\varphi_{abc}\rangle_{T_{1}T_{2}T_{3}}(|\varphi_{a\oplus 1,b,c}\rangle+(-1)^{a}|\varphi_{a,b\oplus 1,c\oplus 1}\rangle)_{\hat{P}_{1}P_{2}P_{3}},
\end{eqnarray}
It's easy to see that after TP's measurements, the particles in three participants' hands will collapse to $|\varphi_{a\oplus 1,b,c}\rangle$ or $|\varphi_{a,b\oplus 1,c\oplus 1}\rangle$. In spite of which state the particles are, $\hat{\rm P}_{1}$ can only perform $\{I, X\}$ operations randomly on the particles. In this way, $\hat{\rm P}_{1}$ will introduce a probability of $\frac{1}{2}$ in random selection, which means that his impersonation can be detected in the second eavesdropping detection.

Suppose that three participants' particles are in the state $|\varphi_{a,b\oplus 1,c\oplus 1}\rangle$ and $\hat{\rm P}_{1}$ selects the operation $I$. Then, three participants respectively perform $Z^{a}\otimes X^{b}\otimes X^{0}$ on $|\varphi_{a,b\oplus 1,c\oplus 1}\rangle$ in Step 6, i.e., $Z^{a}\otimes X^{b}\otimes X^{0}|\varphi_{a,b\oplus 1,c\oplus 1}\rangle=|\varphi_{0,1,c\oplus 1}\rangle$. When P$_{i}$ measures the detecting particles in the basis $B_{x}$, no errors are to occur in the second eavesdropping detection. However, his impersonation will definitely expose when P$_{i}$ measures the detecting particles in the basis $B_{z}$. In spite of successfully evading the first eavesdropping detection, $\hat{\rm P}_{1}$'s attack will be detected with the the probability of $(1-(\frac{5}{8})^\delta)$ in the second eavesdropping detection. Therefore, this protocol can stand against the impersonation attack.

It is worthy noting that the first eavesdropping detection cannot be omitted in this protocol. The classical hash function used in this protocol is not unconditionally safe for its security based on computational complexity. This implies that it might be insecure in quantum computation environment. Therefore, for the safety of $k_{i}$, the value of $t_{i}$ should be kept private. However, the external eavesdropper Eve and the internal dishonest participant $\bar{\rm P}_{1}$ can eavesdrop all or part of the information about $t_{i}$ if there's only the second eavesdropping detection. By measurements on particle sequences and TP's measurement results, attackers may infer the private key $k_{i}$ by the following attack strategy.

In this attack that is similar to intercept-resend attack, Eve intercepts particle sequences $S_{i}^{'}$ and sends the pre-prepared fake particle sequences to P$_{i}$ between Step 2 and 3. After Step 4, Eve can obtain $t_{i}$ by measuring the intercepted particle sequences, which helps her gain $k_{i}$. Although Eve's attack will be detected in Step 7, she still can eavesdrop all or part of the information about $k_{i}$. From the analysis above, it is shown that the first eavesdropping detection is indispensable because it can successfully prevent eavesdropping $k_{i}$. As a consequence, the first and second eavesdropping detection work together to protect the security of the proposed protocol.
\section{Multi-party QKA protocol}

In this section, we generalize the above three-party QKA protocol to a multi-party case. In this multi-party QKA protocol, $m$ participants P$_1$, P$_2$, $\cdots$, P$_m$ wish to jointly negotiate a secret key $K$, the length of which is $n$. Meanwhile, they hope to authenticate the identity of the participants with the assistance of TP. Beforehand, each participant P$_i(i=1,2,\cdots,m)$ has his own identity information ID$_i$ and shares a private key $k_i$ with TP. The detailed process of this protocol is as follows.

\noindent Step 1 \begin{minipage}[t]{0.9\linewidth}TP and participant P$_i(i=1,2,\cdots,m)$ separately generate a random number $r_T$ and $r_i$, and declare them publicly. Then TP randomly selects a hash function $h$ from a hash function cluster and announces it, where $h:2^{*}\rightarrow2^{L}$, and $L=n+m\cdot\delta$. Each participant P$_i$ and TP calculate the hash value $t_i$, where $t_i=h_{k_i}(ID_i||r_i||r_T)$.
\end{minipage}

\noindent Step 2 \begin{minipage}[t]{0.9\linewidth}TP prepares $m$ ordered sequences of $L$ Bell states. Each EPR pair is in the initial state $|\phi^{+}\rangle$. TP uses first qubits of each Bell state to form $m$ ordered sequences $Q_1,Q_2,\cdots,Q_{m}$. Similarly, he forms another $m$ ordered sequences $S_1,S_2,\cdots,S_{m}$ with all the second qubits. Here, $H_{i}$, $S_{i}(i=1,2,\cdots,m)$ denote the first and second particles of the Bell state $|\phi^{+}\rangle$. Then, TP generates $m\zeta$ decoy particles which are randomly in four states $\{|0\rangle,|1\rangle,|+\rangle,|-\rangle\}$. He respectively inserts $\zeta$ decoy particles randomly in sequences $S_1,S_2,\cdots,S_{m}$ and obtains $m$ new particle sequences $S_1^{'},S_2^{'},\cdots,S_{m}^{'}$. Finally, TP sends $S_1^{'},S_2^{'},\cdots,S_m^{'}$ to P$_1$, P$_2$, $\cdots$, P$_m$ respectively.
\end{minipage}

\noindent Step 3 \begin{minipage}[t]{0.9\linewidth}(The first eavesdropping detection) Confirming that P$_i$ has received the particle sequence $S_i^{'}$, they begin to check the eavesdrop. For sequence $S_{i}^{'}$, TP announces the positions and the corresponding basis ($B_{z}$ or $B_{x}$) of decoy particles. Then P$_i$ measures the decoy particles in correct basis and randomly announces half of the measurement results. TP announces the initial states of the left half of decoy particles. At last, they check whether the initial states and the measurement results are consistent. If they are consistent, TP and P$_i$ think the quantum channel between them is secure; otherwise, they abandon the protocol.
\end{minipage}

\noindent Step 4 \begin{minipage}[t]{0.9\linewidth}Both P$_i$ and TP perform operation $\{I,H\}$ respectively on two particles of one Bell state according to the value of $t_i$. Concretely, if $t_i=0$, they perform operation $I$; otherwise, they perform operation $H$. Obviously, if the selected operations of P$_i$ and TP are the same, the state of each EPR pairs stays unchanged. Otherwise, the Bell states will be in the form as Eq.(2).
\end{minipage}

\noindent Step 5 \begin{minipage}[t]{0.9\linewidth}TP sequentially makes GHZ states measurements on $m$ particles in his hand, and announces the measurement results $|\varphi_{u_{1},u_{2},\cdots,u_{m}}\rangle$. Similar to Eq.(6), the corresponding particles in three participants' hands will collapse to $|\varphi_{u_{1},u_{2},\cdots,u_{m}}\rangle$.
\end{minipage}

\noindent Step 6 \begin{minipage}[t]{0.9\linewidth}P$_i$ selects an appropriate operation from $\{I,X,Z\}$ according to the result announced by TP. Specifically, when the result is $|\varphi_{u_{1},u_{2},\cdots,u_{m}}\rangle$, P$_1$, P$_2$,$\cdots$,P$_m$ performs $Z^{u_{1}},X^{u_{2}},\cdots,X^{u_{m}}$ on their particles respectively, where $Z^{u_{1}}\otimes X^{u_{2}}\otimes\cdots\otimes X^{u_{m}}|\varphi_{u_{1},u_{2},\cdots,u_{m}}\rangle=|\varphi_{00\cdots0}\rangle$. In this way, three participants can share $L$ GHZ entangled states, all of which are in the state $|\varphi_{00\cdots0}\rangle$.
\end{minipage}

\noindent Step 7 \begin{minipage}[t]{0.9\linewidth}(The second eavesdropping detection) P$_1$ randomly selects $\delta$ particles from the particle sequence $S_{1}$ for eavesdropping detection, and informs the other $m-1$ participants P$_2$, P$_3$, $\cdots$, P$_m$ of the positions of detecting particles. For each detecting particle, P$_1$ randomly measures it in basis $B_{z}$ or $B_{x}$. Then, he requires P$_2$, P$_3$, $\cdots$, P$_m$ to measure their qubits in the same basis as his, respectively. After that, P$_2$, P$_3$, $\cdots$, P$_m$ tell P$_1$ their measurement outcomes. Suppose that the measurement outcomes of $m$ participants in the basis $B_{z}$ are $j_{P_i}^z=\{+1,-1\}$, and that in $B_{x}$ are $j_{P_i}^x=\{+1,-1\}$. According to
\begin{eqnarray}
|\varphi_{00\cdots0}\rangle&=&\frac{1}{\sqrt{2}}(|0\rangle|0\rangle\cdots|0\rangle+|1\rangle|1\rangle\cdots|1\rangle)
\nonumber\\
&=&\frac{1}{2^{\frac{m+1}{2}}}\sum_{\tau_{1}\tau_{2}\cdots\tau_{m}}(|\tau_{1}\tau_{2}\cdots\tau_{m}\rangle),
\end{eqnarray}
where $\tau_{i}\in\{+,-\}$($i=1,2,\cdots,m$), and $\tau_{1}\oplus \tau_{2}\oplus \cdots \oplus \tau_{m}=0$. P$_{1}$ checks whether or not $j_{P_1}^z=j_{P_2}^z=\cdots=j_{P_m}^z$ is satisfied for those measurements in $B_{z}$ (whether $\prod_{i=1}^{m}j_{P_i}^x=1$ is satisfied or not in $B_{x}$). If the error rate exceeds the threshold, they abandon the protocol. Otherwise, they continue the protocol.
\end{minipage}

\noindent Step 8 \begin{minipage}[t]{0.9\linewidth} The rest participants P$_2$, P$_3$, $\cdots$, P$_m$ execute the similar eavesdropping detection as P$_1$ does in Step 7. Obviously, the detecting particles that P$_2$, P$_3$, $\cdots$, P$_m$ select do not have a common element with that of P$_1$.
\end{minipage}

\noindent Step 9 \begin{minipage}[t]{0.9\linewidth}Each participant measures his remaining particles in the basis $B_{z}$ and can get the raw agreement key $K$.
\end{minipage}

\section{Conclusion}

In the practical implementation of key agreement, how to authenticate the identity of each participant is a key issue and usually ignored in designing quantum key agreement protocols. In response to this problem, a multi-party QKA protocol with authentication is proposed. In this protocol, Bell states are used as information carriers, and generated by a semi-trusted third party who helps the participants achieve key agreement task. Classical hash function and Hadamard operation are combined to authenticate the identity of participants. It is worth emphasizing that although the classical hash function is introduced, it does not reduce the security of the protocol. This is because the hash value in the protocol is not declared publicly. In this case, each participant's private key is still secure even if the hash function is broken by quantum computing. This implies that these private keys can be reused, which also greatly improves the practicability of the proposed protocol. Furthermore, based on decoy-state method and correlations of GHZ states, we design two eavesdropping detection processes, which make the protocol stand against some common external/internal attacks and the impersonation attack. The security analysis shows that the proposed protocol satisfies the requirements of fairness and security.

\section*{Acknowledgments}

\end{document}